\documentclass[prb,twocolumn,showpacs,floatfix]{revtex4}
\usepackage{graphicx}
\usepackage{dcolumn}
\usepackage{bm}
\begin{document}
\title{Magnetic ordering in double perovskites  R$_2$CoMnO$_6$ (R= Y,Tb) investigated by high resolution neutron spectroscopy}
\author{Tapan Chatterji$^1$, Bernhard Frick$^1$ and Harikrishnan S. Nair$^2$}
\address{$^1$Institut Laue-Langevin, BP 156, 38042 Grenoble Cedex 9, France\\
$^2$Forschungszentrum J\"ulich, J\"ulich, Germany
}
\date{\today}

\begin{abstract}
We have investigated low energy nuclear spin excitations in double perovskite compounds R$_2$CoMnO$_6$ (R = Y,Tb) by inelastic neutron scattering with a high-resolution back-scattering spectrometer. We observed inelastic signals at about 2.1 $\mu$eV for Y$_2$CoMnO$_6$ and also for Tb$_2$CoMnO$_6$ at T = 2 K in both energy loss and energy gain sides. We interpret these inelastic peaks to be due to the transitions between the hyperfine split nuclear levels of $^{59}$Co nucleus. The inelastic peaks move towards the central elastic peak and finally merge with it at the magnetic ordering temperature $T_C$. The energy of the low energy excitations decreases continuously and becomes zero at $T_C \approx 75$ K for Y$_2$CoMnO$_6$ and $T_C \approx 100$ K for Tb$_2$CoMnO$_6$.  For Tb$_2$CoMnO$_6$, which contains magnetic rare-earth ions, additional quasielastic scattering  due presumably to the fluctuations of large Tb magnetic moments was observed. The present study reveals  the magnetic ordering of the Co sublattice. The results of this investigation along with that obtained by us for other compounds indicate the presence of unquenched orbital moments in some of the Co compounds.
\end{abstract}
\maketitle
\section{Introduction}
Rare earth double perovskites R$_{2}$BB$^\prime$O$_6$ where R is Y or a rare-earth element and  B and B$^\prime$ are transition elements,  are strongly correlated electron compounds that have drawn a lot of interest lately. Initially these compounds were investigated \cite{wold58,goodenough61} for the verification of the Goodenough-Kanamori rules \cite{goodenough55,kanamori59}, which predict that ferromagnetism results when an empty d orbital of one metal site interacts with a half-filled d-orbitals of another metal site through an anion in a 180$^\circ$ superexchange interaction. Recently these compounds have been studied for their possible applications as high temperature ferromagnetic semiconductors in spintronics. These compounds have been found to show large magnetocapacitance,
cationic-ordering and are even predicted to exhibit polar behaviour \cite{rogado05}. They crystallize in either
monoclinic $P21/n$ space group or in orthorhombic $Pnma$. In the monoclinic space group, symmetry allows
the B cations to occupy the Wyckoff positions 2c or 2d and hence, a B-site ordered structure can result \cite{bull03}.
The cationic ordering at the B-site is an important issue which has direct correlation to magnetism and
electrical transport in double perovskites (DP) in general\cite{ogale99}. The ferromagnetic properties of DPs were
explained \cite{goodenough55} based on double exchange between B$^{2+}$ - O - B$^{\prime 4+}$ . But perfect ordering is never attained and the
mixed occupation of the B-site by the two cations, known as antisite disorder plays an important and
influential factor in the magnetic properties of these materials. The antisite disorder can lead to
superexchange interactions between the B$^{2+}$ and B$^{\prime 4+}$ cations, as does stabilization of other valence of B (like
B$^{3+}$ or B$^{\prime'3+}$). Even though La-based double perovskites with B/B$^{\prime}$ occupied by Co, Mn, Ni etc have received
much attention in this regard, R$_2$BB$^\prime$O$_6$ with small-radius, magnetic rare earth have rarely been reported\cite{booth09,truong11,sazanov07}.
The small-radii-rare-earth DPs are interesting because the small radius of R is propitious for the ordered
cationic arrangement at the B/B$^\prime$ sites in addition to the structural distortions that can result from ionic radii
difference. Apart from size effects and crystallographic ordering, the spin states of the B-site ions are also
important, for example, Co can assume different valence and spin states in these materials. Another
important parameter is the magnetocrystalline anisotropy of the rare earth ion that can show up as
interesting magnetic properties at low temperatures. Thus it is clear that a correct explanation of the
observed macroscopic properties of DPs can be advanced only after estimating the valence, spin states and a
thorough crystallographic characterization which consequently quantifies the anti-site disorder of the B-site
ions. 

Here we have investigated the magnetic ordering in Y$_2$CoMnO$_6$ and Tb$_2$CoMnO$_6$ by a less-known technique of high resolution neutron spectroscopy. This technique, described in detail in section \ref{hyperfine}, enables one to probe the magnetic ordering of Co ions,  thanks to the large spin dependent scattering cross section of the $^{59}$Co isotope which has 100\% natural abundance. By proper calibration one can also estimate the ordered magnetic moment of Co ions in these double perovskite compounds. Neutron diffraction on the contrary can only determine the average magnetic moment of Co and Mn in these double perovskite compounds. 

\section{Synthesis and characterization}
\begin{figure}
\resizebox{0.35\textwidth}{!}{\includegraphics{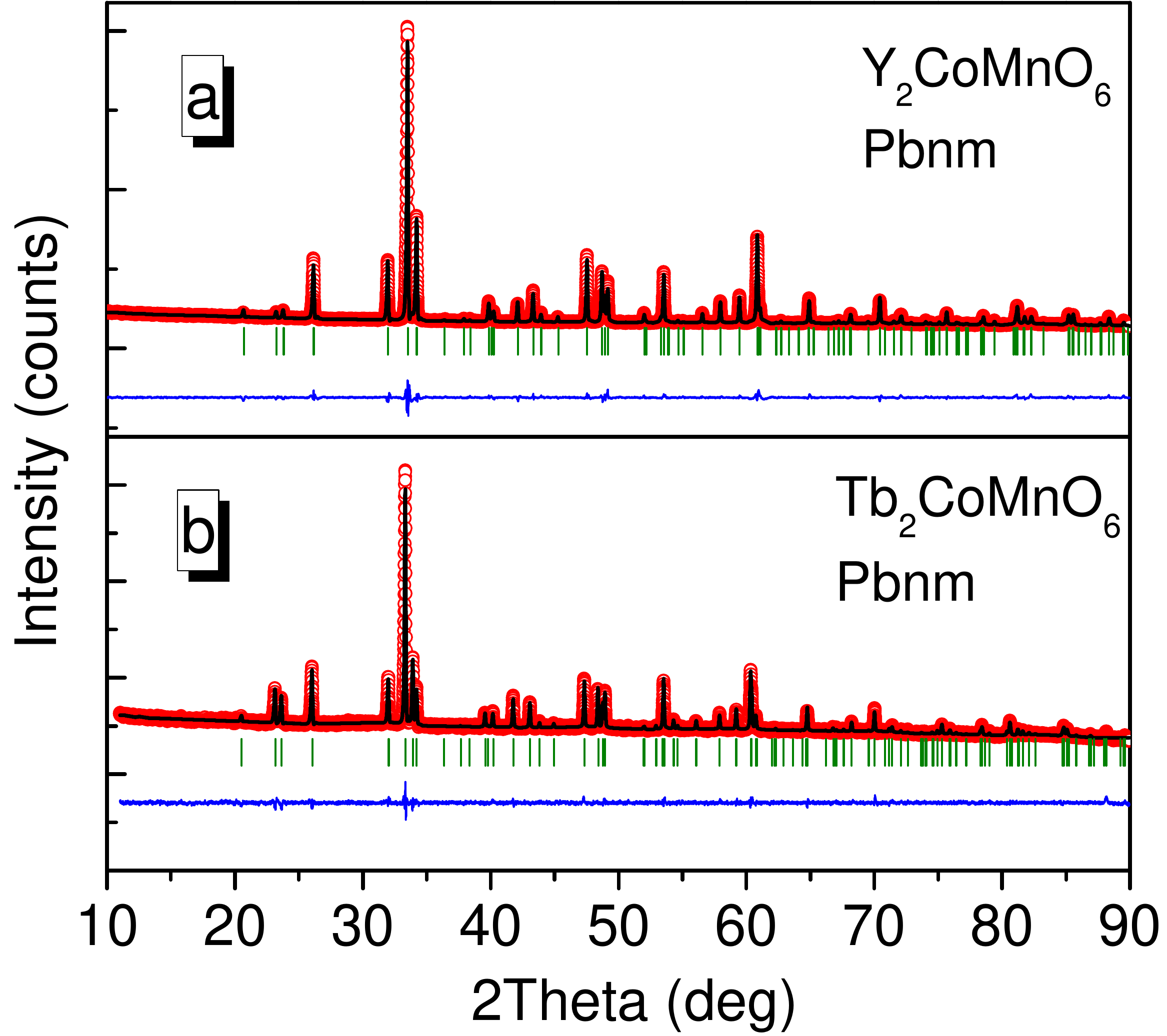}}
\caption {(Color online) Results of the Rietveld refinement of the X-ray diffraction data from (a)   Y$_2$CoMnO$_6$ and (b)   Tb$_2$CoMnO$_6$.          } 
\label{xrayref}
\end{figure}

Polycrystalline powders of Y$_2$CoMnO$_6$ and Tb$_2$CoMnO$_6$ were prepared through conventional
solid state reaction using 4N purity Y$_2$O$_3$, Tb$_2$O$_3$, Co$_3$O$_4$ and MnO$_2$. After mixing the chemicals
in desired stoichiometric ratios, they were ground using a mortar and pestle and heat
treated at 1320 C for 48 h. The cycle of grinding and heating was repeated till a
homogeneous phase was obtained. Room temperature powder x-ray diffraction
experiments were performed on a Huber Diffractometer with Guinier geometry using Cu K$\alpha$ radiation. The results of refined lattice parameters in the monoclinic space group  $P2_1/n$ are $a = 5.2330(3)$,
                             $b = 5.5929(3)$,
                             $c = 7.4692(4)$ {\AA},
                             $\beta = 89.954(7) ^\circ $ for Y$_2$CoMnO$_6$ with the 
reliability factors:  $R_p = 0.149$,     $R_{wp} =  0.10$  and   $\chi ^2  =  4.09$. The corresponding results for Tb$_2$CoMnO$_6$ are $a = 5.2777(5)$,
                              $b = 5.5839(5)$,
                              $c = 7.5119(7)$ {\AA},
                              $\beta = 90.009(4)^\circ $ and the
reliability factors are  $R_p = 0.301$,     $R_{wp} =  0.155$,     $\chi^2 = 1.11$. The X-ray data can also be refined in orthorhombic space group $Pnma$. The refinement in  $Pbnm$ gives the following results for Y$_2$CoMnO$_6$:
$a = 5.2331(3)$,
$b = 5.5929(4)$,
$c = 7.4693(5)$ {\AA}, the unit cell volume
$V= 218.621(2) {\AA}^3$,
$R_{Bragg} = 0.249,
R_p = 0.156; R_{wp} = 0.108; \chi^2 = 4.8$. For 
Tb$_2$C0MnO$_6$ the refinement in the orthorhombic space group $Pbnm$ gives
$a = 5.2795(5),
b = 5.5860(5),
c = 7.5145(7)$ {\AA}, the unit cell volume 
$V = 221.617(4) {\AA}^3$,
$R_{Bragg} = 0.0358,
R_p = 0.288; R_{wp} = 0.149; \chi^2 = 1.2$.
Fig. \ref{xrayref} shows the Rietveld refinement of the X-ray diffraction data in the orthorhombic space group  $Pbnm$. 
With the present data it is not easy to distinguish between the monoclinic
and the orthorhombic space groups.

\section{Magnetic measurements}
\begin{figure}
\resizebox{0.35\textwidth}{!}{\includegraphics{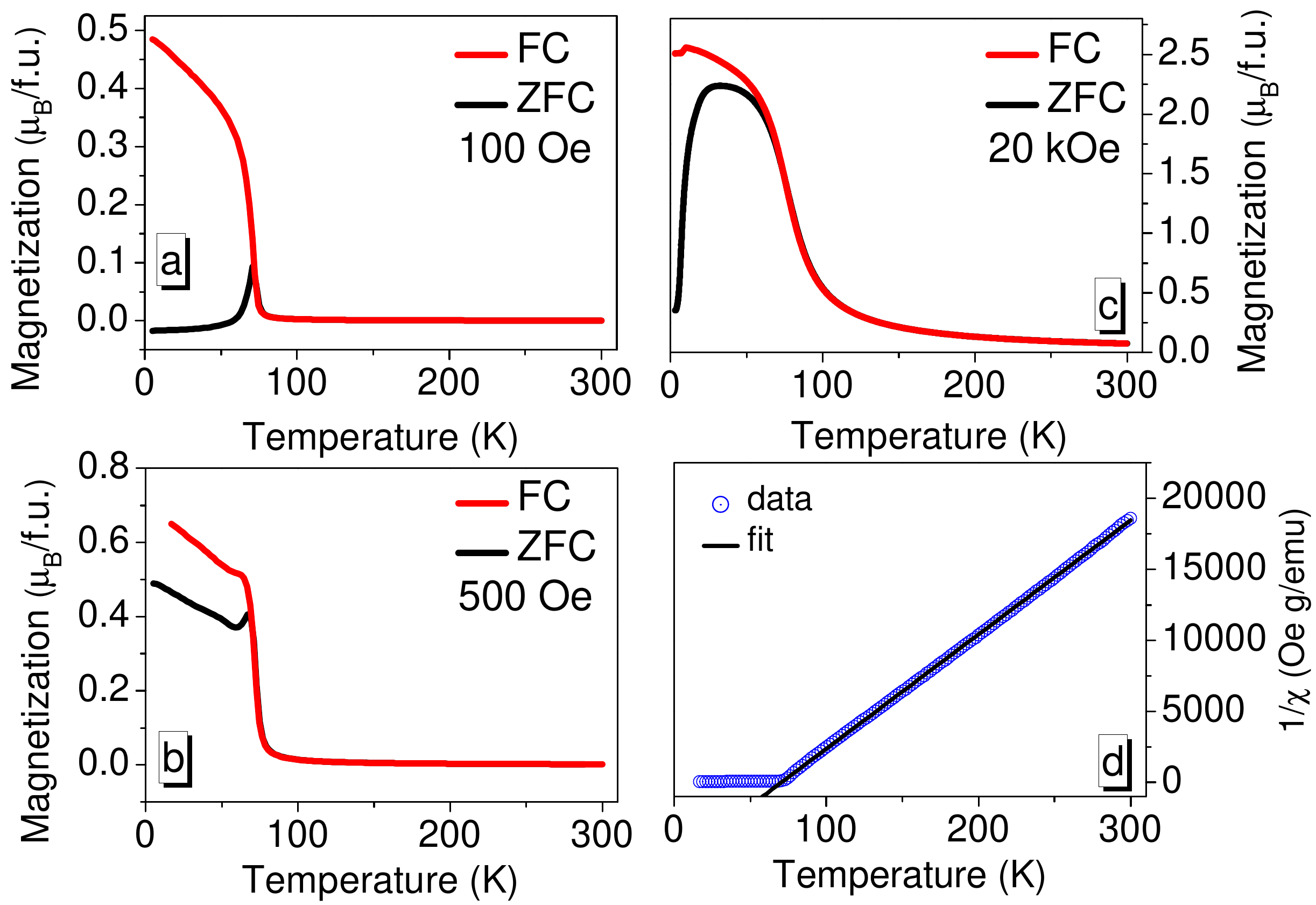}}
\caption {(Color online)  Magnetization profiles for Y$_2$CoMnO$_6$ in zero-field-cooled (ZFC) and field-cooled (FC) cycles with applied fields of (a) 100 Oe, (b) 500 Oe, and (c) 20 kOe. The phase transition from the paramagnetic to ferromagnetic phase occurs at $T_C \approx 70$ K. (d) shows the Curie-Wiess fit to inverse magnetic susceptibility.        } 
\label{YCMO_mag}
\end{figure}

\begin{figure}
\resizebox{0.35\textwidth}{!}{\includegraphics{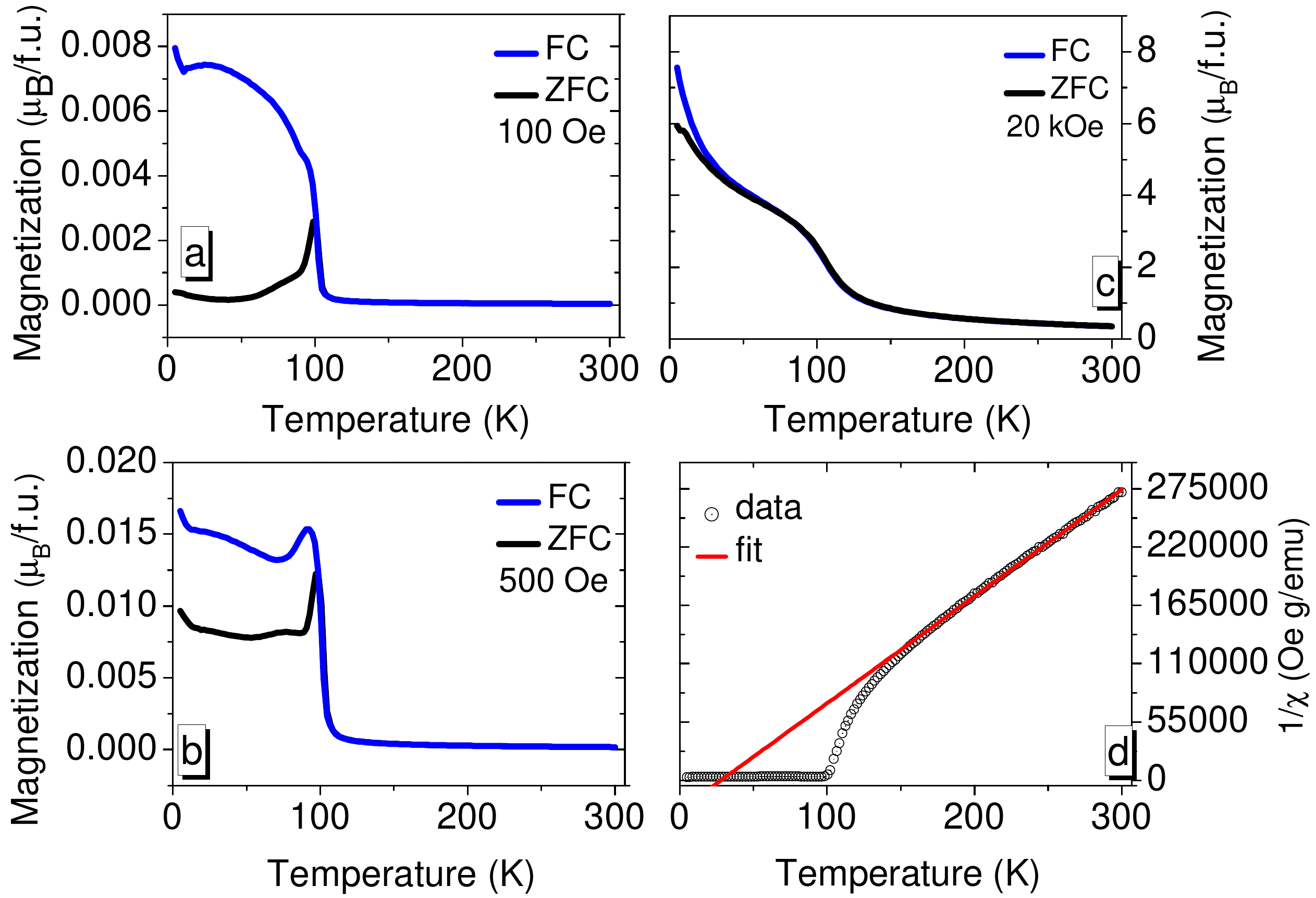}}
\caption {(Color online)  Magnetization profiles for Tb$_2$CoMnO$_6$ in zero-field-cooled (ZFC) and field-cooled (FC) cycles with applied fields of (a) 100 Oe, (b) 500 Oe, and (c) 20 kOe. The phase transition from the paramagnetic to ferromagnetic phase occurs at $T_C \approx 98$ K. (d) shows the Curie-Wiess fit to inverse magnetic susceptibility.        } 
\label{TCMO_mag}
\end{figure}

Magnetic measurements were performed using a SQUID magnetometer as well as Physical
Property Measurement System (both Quantum Design). Fig. \ref{YCMO_mag} shows the results of magnetization measurements on Y$_2$CoMnO$_6$. Fig. \ref{TCMO_mag} shows the results of magnetization measurements on Tb$_2$CoMnO$_6$. The magnetic transition temperatures are determined by measuring temperature evolution of magnetization.
The T$_c$ for Y$_2$CoMnO$_6$ is about 70 K and that of Tb$_2$CoMnO$_6$ is about 98 K.

\section{Hyperfine interaction}
\label{hyperfine}
The hyperfine field of an atom or ion is the magnetic field at the atomic nucleus produced by the electrons in the solid due to the hyperfine interaction between the magnetic moment of the electrons and that of the nucleus \cite{freeman65}. This interaction can be measured by the M\"ossbauer effect or by the nuclear magnetic resonance (NMR) technique. Another less well-known method is the spin-flip scattering of neutrons measured by high resolution neutron spectroscopy \cite{heidemann70}. It is well-known that the magnetic hyperfine fields in a solid give valuable information about the electronic structure and the magnetic properties of the solid. The hyperfine field is a valuable probe of electron spin density distribution at the nuclei. It can sometimes be related to the electronic magnetic moment.  The hyperfine field is site and element selective. The hyperfine field $B_{hf}$ can be given by
\begin{equation}
B_{hf}=B^s_{hf}+B^d_{hf}+B^o_{hf}
\end{equation}
where $B^s_{hf}$ is the Fermi contact term due to the s electrons, $B^d_{hf}$ is magnetic dipole and $B^o_{hf}$ is the orbital term due to the non-s electrons. Normally the Fermi contact term is the most dominant term whereas the magnetic dipole term is often very small and can be neglected. The orbital term is appreciable for rare earth ions except for Eu$^{2+}$ and Gd$^{3+}$, which have no orbital moment. The orbital moment is usually quenched in some 3d Fe series elements. However in some compounds of Co and V it can be quite important.

The method of investigating hyperfine interaction by high resolution back-scattering neutron spectroscopy was developed by Heidemann \cite{heidemann70}. Heidemann \cite{heidemann70} worked out 
the double differential cross section of this scattering process. The 
process can be summarized as follows: If  neutrons with spin ${\bf s}$ 
are scattered from  nuclei with spins  ${\bf I}$, the probability that 
their spins will be flipped is $2/3$. The nucleus at which the 
neutron is scattered with a spin-flip, changes its magnetic quantum 
number $M$ to $M\pm 1$ due to the conservation of the 
angular momentum. If the nuclear ground state is split up into 
different energy levels $E_{M}$ due to the 
hyperfine magnetic field or an electric quadrupole interaction, then 
the neutron spin-flip produces a change of the ground state energy 
$\Delta E = E_{M} - E_{M\pm 1}$. This energy change is transferred 
to the scattered neutron. The double differential scattering cross section \cite{heidemann70} is 
given by  the following expressions:
\begin{equation}
	 \left(\frac{d^2\sigma}{d\Omega d\omega}\right)_
{inc}^{0}=\overline{(\bar{\alpha^{2}}-{\bar{\alpha}}^{2}+
\frac{1}{3}{\alpha^{\prime}}^{2}I(I+1))}e^{-2W(Q)}
\delta(\hbar\omega),
\label{heidemann01}
\end{equation}
\begin{equation}
\left(\frac{d^2\sigma}{d\Omega d\omega}\right)_
{inc}^{\pm}=
\frac{1}{3}\overline{{\alpha^{\prime}}^{2}I(I+1)}\sqrt{1\pm\frac{\Delta E}{E_{0}}}e^{-2W(Q)}
\delta(\hbar\omega\mp \Delta E)
\label{heidemann02}	 
\end{equation}
where $\alpha$ and $\alpha^{\prime}$ are coherent and spin-incoherent 
scattering lengths, $W(Q)$ is the Debye-Waller factor and $E_{0}$ is 
the incident neutron energy, $\delta$ is the Dirac delta function.  
If the sample contains one type of isotope then 
$\bar{\alpha^{2}}-{\bar{\alpha}}^{2}$ is zero. Also 
$\sqrt{1\pm\frac{\Delta E}{E_{0}}} \approx 1$ because $\Delta E$ is 
usually much less than the incident neutron energy $E_{0}$. In this case 2/3 of 
incoherent scattering will be spin-flip scattering. 
Also one expects a central elastic peak and two inelastic peaks of 
approximately equal intensities. The $^{59}$Co is such a case. 

\begin{figure}
\resizebox{0.35\textwidth}{!}{\includegraphics{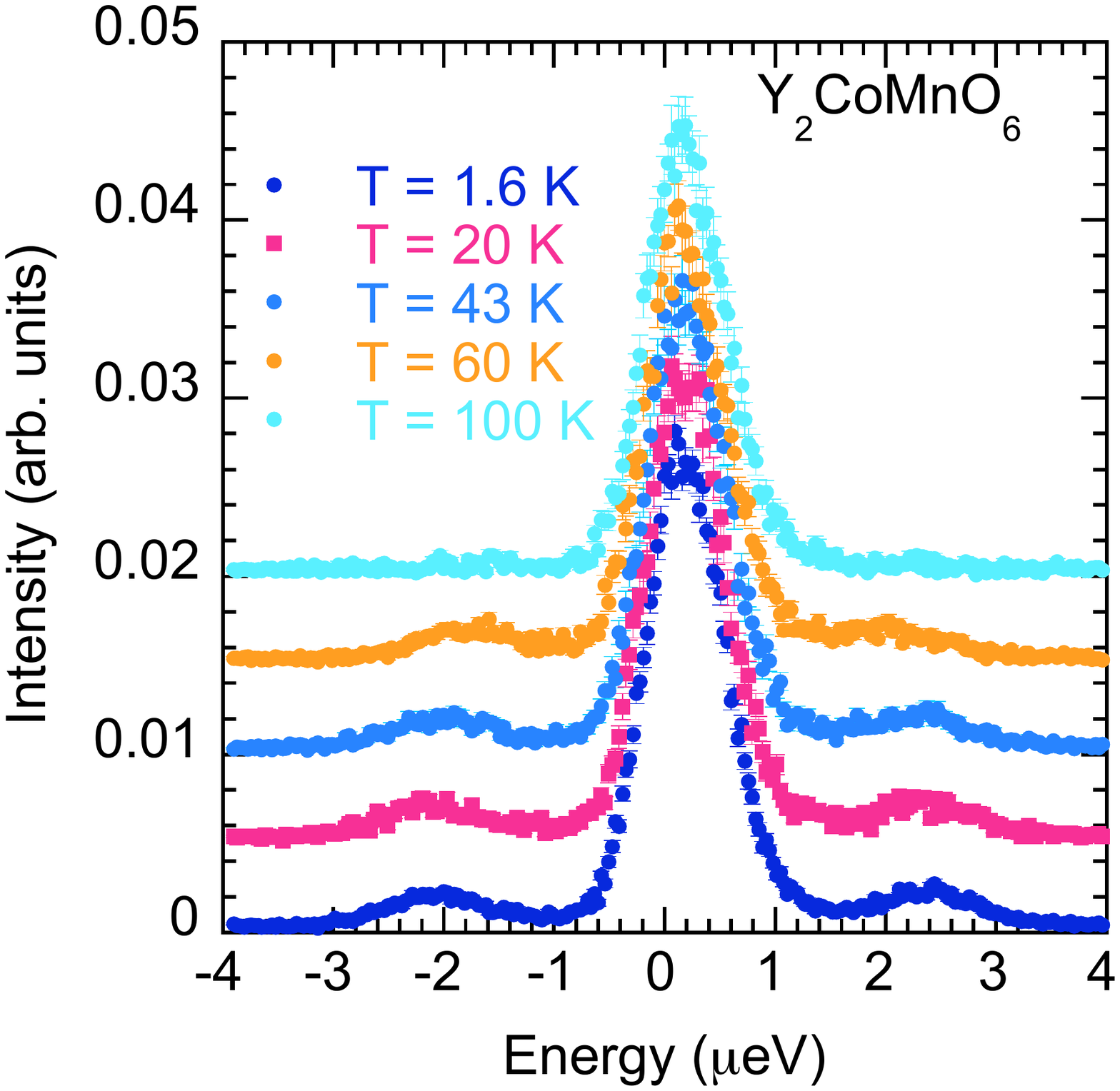}}
\caption {(Color online) Inelastic neutron scattering spectra of Y$_2$CoMnO$_6$ at several temperatures.              } 
\label{structure}
\end{figure}

\begin{figure}
\resizebox{0.35\textwidth}{!}{\includegraphics{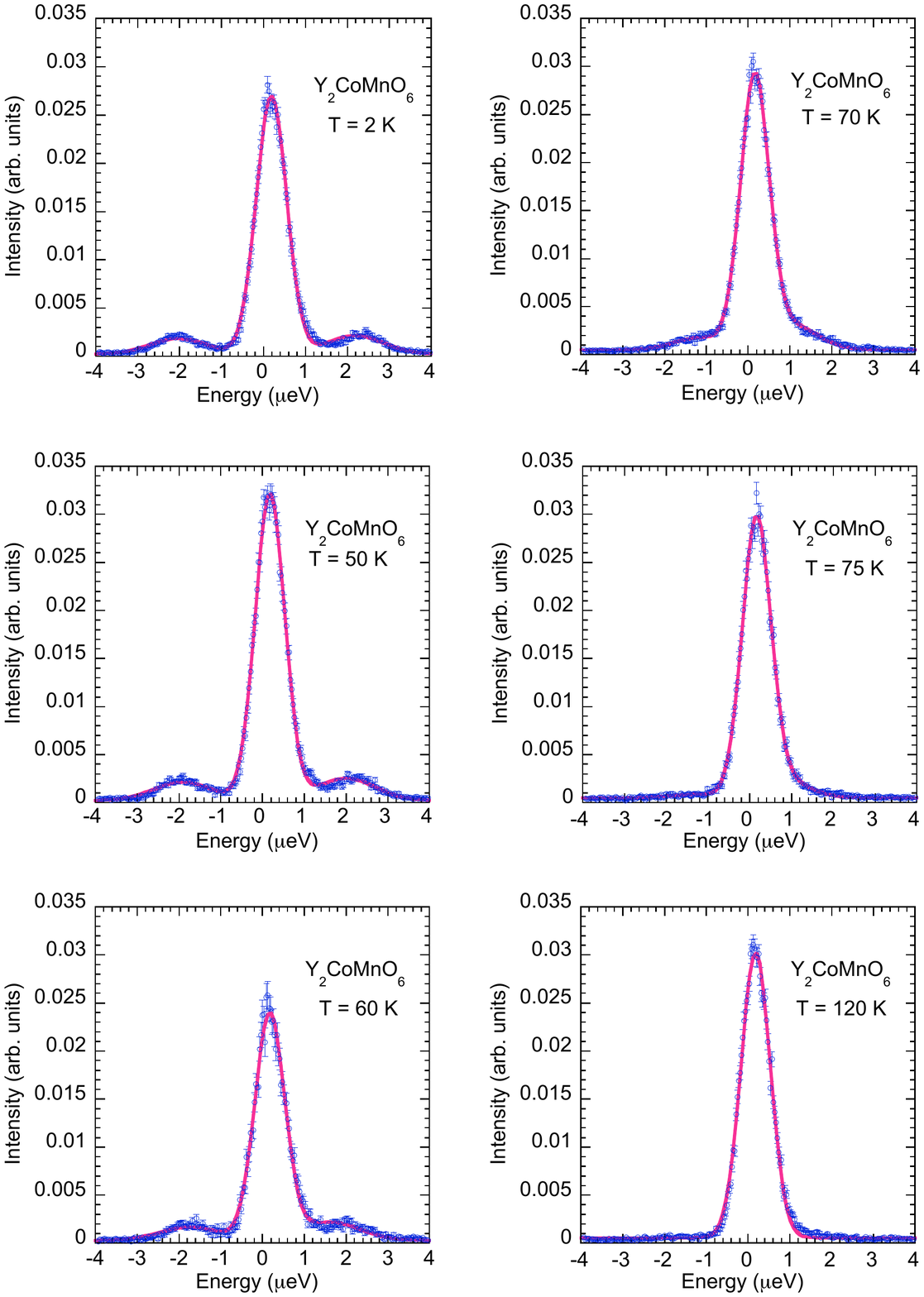}}
\caption {(Color online) Energy spectra of Y$_2$CoMnO$_6$ at several temperatures. The continuous curves are fits of the elastic and the two inelastic peaks with three Gaussian functions.
            } 
\label{spectra}
\end{figure}

\begin{figure}
\resizebox{0.4\textwidth}{!}{\includegraphics{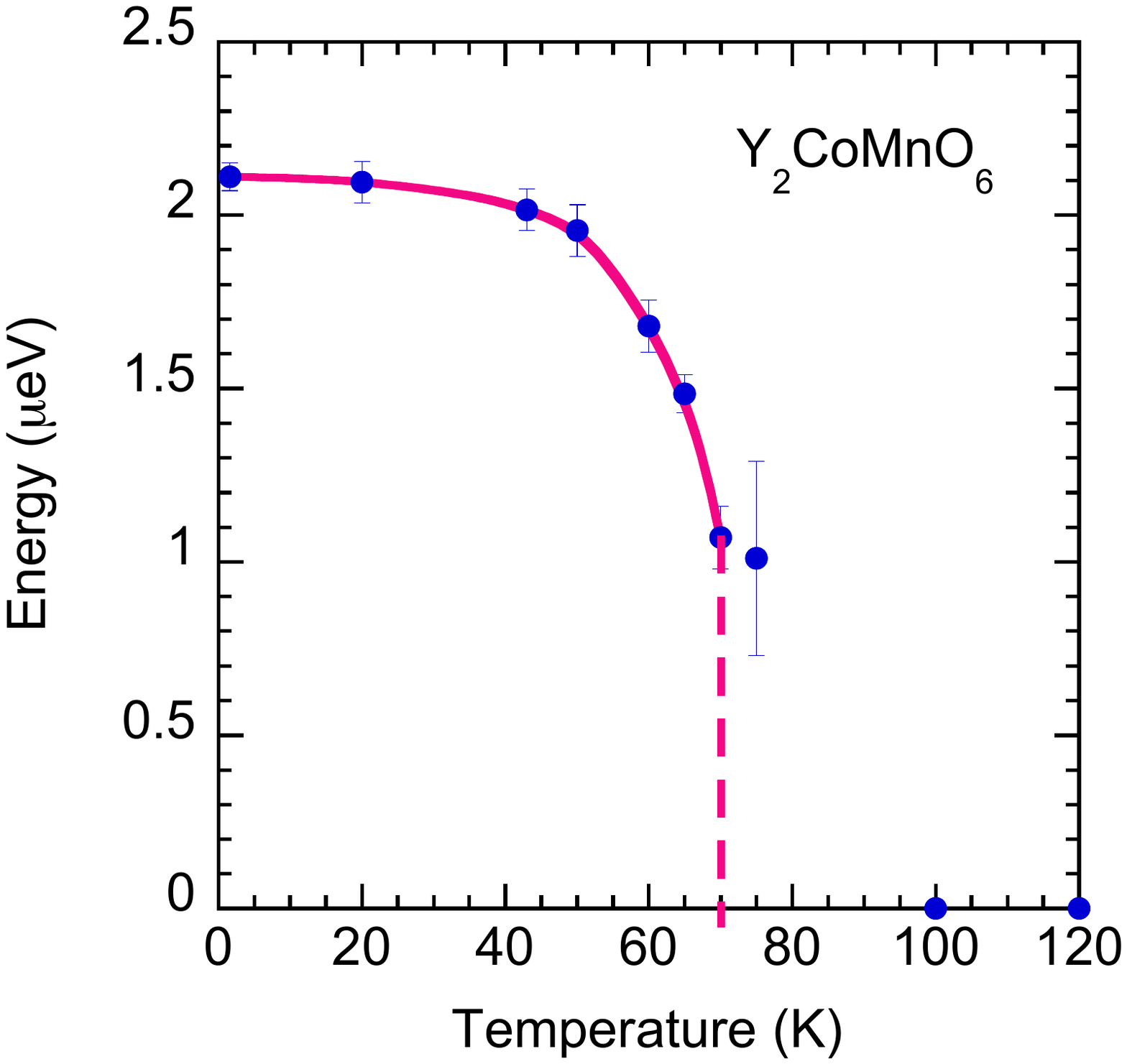}}
\caption {(Color online) Temperature variation of the energy of the inelastic peak 
          of Y$_2$CoMnO$_6$ . }
\label{E(T)}
\end{figure}
\begin{figure}
\resizebox{0.35\textwidth}{!}{\includegraphics{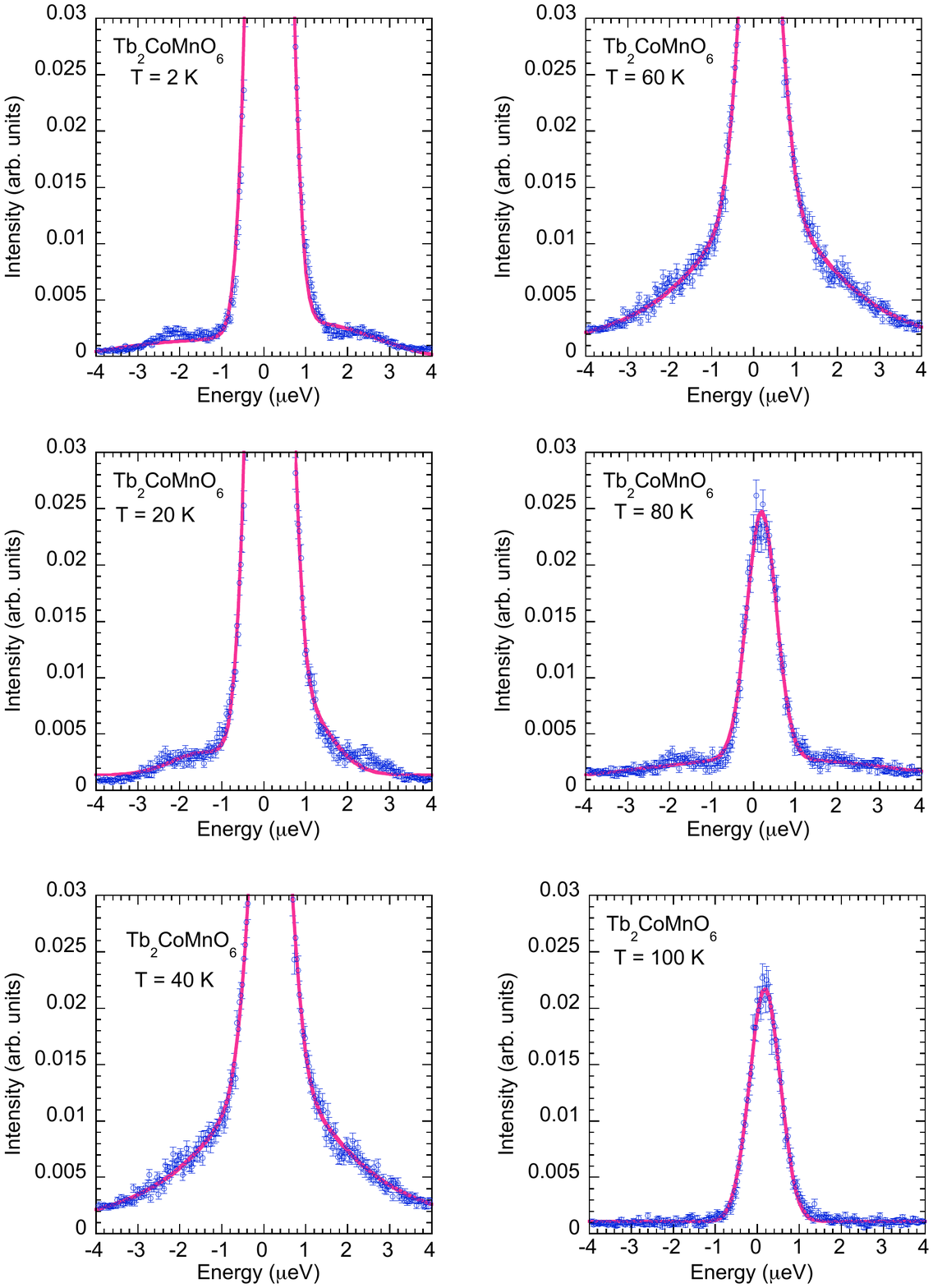}}
\caption {(Color online)    Energy spectra of Tb$_2$CoMnO$_6$ at several temperatures.   The red continuous curves are attempts to fit the data. At low temperatures T = 2 and 20 K and also at T = 80 K where the quasielstic scattering is very broad and mostly outside of the energy window of the instrument, three Gaussians peaks have been fitted, but the results are unsatisfactory. At T = 40 and 60 K a Gaussian peak for the elastic peak and a Lorentzian for the quasielastic scattering have been used to fit the data and the fits were satisfactory.  At T = 100 K only a Gaussian function has been fitted.      } 
\label{tbspectra}
\end{figure}

 We investigated previously hyperfine interaction in several Nd compounds  \cite{chatterji00,chatterji02,chatterji04,chatterji04a,chatterji08,chatterji08a,chatterji09f} by high resolution neutron spectroscopy and found that the hyperfine splitting of the Nd nuclear levels is linearly proportional to the ordered electronic magnetic moment of Nd. Our recent investigation on a series of Co compounds \cite{chatterji09c,chatterji09d,chatterji10,chatterji11} showed that this simple relationship is no longer valid for Co compounds presumably due to the unquenched orbital moments in Co-compounds. It is known that in Co metal and also in some Co-compounds the sign of the hyperfine field due the electronic orbital magnetic moment is opposite to that due to the spin moment. The orbital moments in different Co-compounds are different and often unknown. The determination of orbital moment is not easy and involves either polarized neutron diffraction or X-ray magnetic scattering or x-ray magnetic circular dichroism (XRMD) techniques. What we usually know is the total ordered magnetic moment by unpolarised neutron diffraction. So the study of hyperfine interaction may also give useful information about the orbital magnetic moment. We therefore studied a series of Co compounds \cite{chatterji09c,chatterji09d,chatterji10,chatterji11,chatterji10a,chatterji11a} by  high-resolution back-scattering neutron spectroscopy. Also Heidemann \cite{heidemann75a} and Heidemann et al. \cite{heidemann75b} studied Co and Co-P amorphous alloys and La-Co, Y-Co and Th-Co intermetallic compounds. Here 
we report the results of the studying  the hyperfine interaction in the double perovskites  Y$_2$CoMnO$_6$ and Tb$_2$CoMnO$_6$ . 

 \section{High resolution inelastic neutron scattering investigations}
  
  \begin{table}[ht]
\caption{Ordered electronic moment of Co and the energy of Co nuclear spin excitations }
\label{table1}
\begin{center}
\begin{tabular}{lccc} \hline \hline
\emph{Compound} & Moment ($\mu_B$)&\emph{$\Delta E (\mu eV)$}  & \emph{Reference}\\ \hline
Y$_2$CoMnO$_6$& - & 2.11(4)&[present work]\\
Tb$_2$CoMnO$_6$& - & 2.1(1)&[present work]\\
CoV$_2$O$_6$& 3.5(1)& 1.38(5)&[23]\\
CoCl$_2$& 3.0(1)& 1.34(3)&[25]\\
Co$_2$SiO$_4$& 3.61(3)& 1.387(6)&[22]\\
CoF$_2$ & 2.60(4)&0.728(8)&[21]\\
CoO & 3.80(6)&2.05(1)&[20]\\
Co & 1.71 & 0.892(4)&[24,26]\\
Co$_{0.873}$P$_{0.127}$ & 1.35&0.67&[26]\\
Co$_{0.837}$P$_{0.161}$ &1.0& 0.54&[26]\\
Co$_{0.827}$P$_{0.173}$& 1.07&0.56&[26] \\
Co$_{0.82}$P$_{0.18}$ & 0.93&0.49 &[26]\\ 
LaCo$_{13}$&1.58&0.69&[27]\\
LaCo$_{5}$&1.46&0.32&[27]\\
YCo$_{5}$&1.51&0.37&[27]\\
ThCo$_{5}$&1.02&0.31&[27]\\\hline
\end{tabular}
\end{center}
\end{table}
\begin{figure}
\resizebox{0.4\textwidth}{!}{\includegraphics{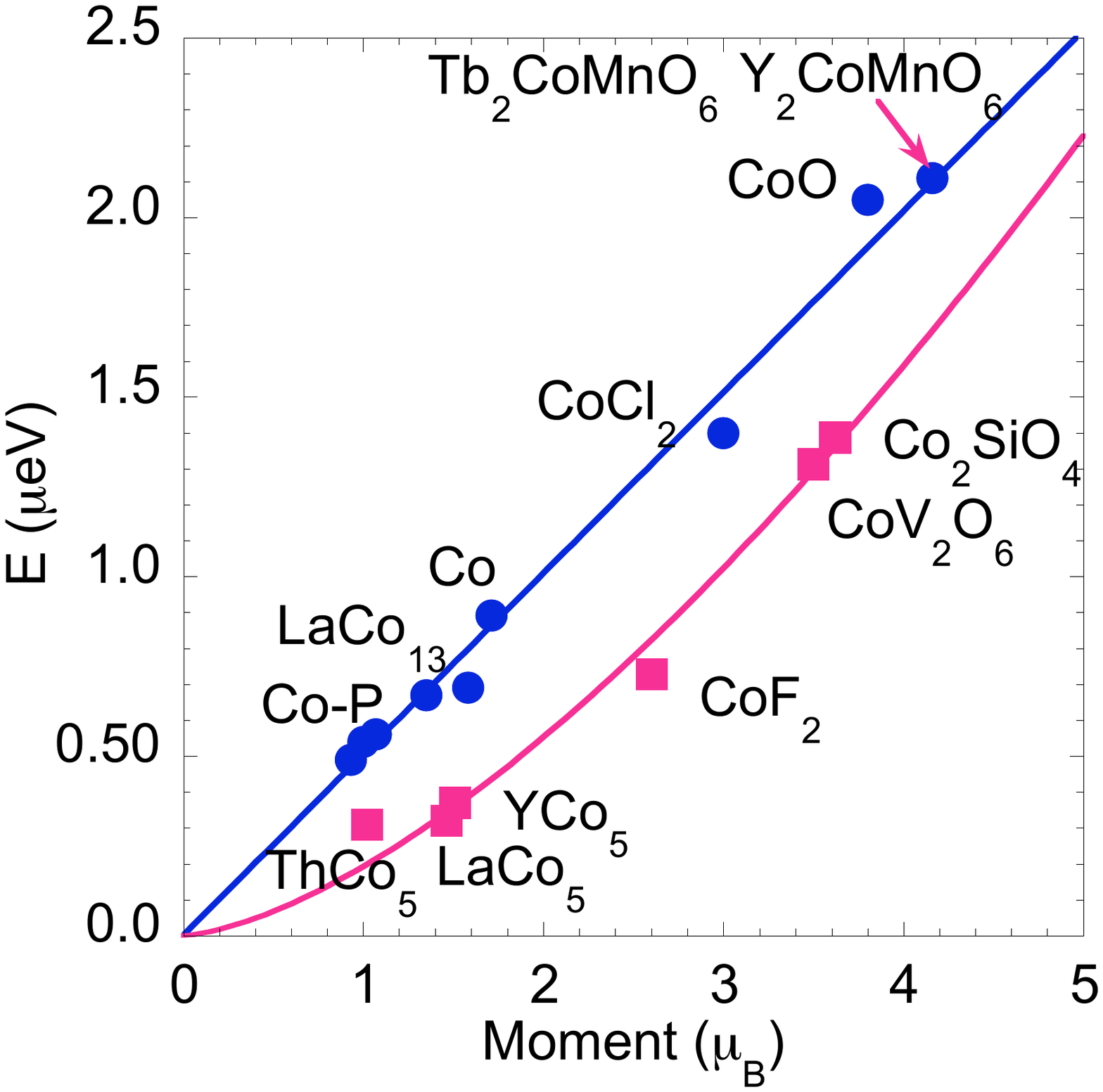}}
\caption {(Color online) Plot of the energy of inelastic signal vs. ordered electronic moment of Co-based materials. The continuous curves are linear and power law fit of the data corresponding to the normal and anomalous compounds.}
\label{cocompounds}
\end{figure}

We performed inelastic neutron scattering experiment on the back-scattering neutron spectrometer IN16 of the Institute Laue-Langevin. The neutron wavelength was 6.271 {\AA}. About 5 g of powder  Y$_2$CoMnO$_6$ and Tb$_2$CoMnO$_6$  samples were placed inside a flat Al sample holder that was fixed to the cold tip of the standard He cryofurnace. 

Fig. \ref{spectra} shows energy spectra obtained from  Y$_2$CoMnO$_6$  at several temperatures. We checked carefully the Q dependence of the inelastic peaks and found that they were Q independent as expected for hyperfine peaks. So we integrated over all measured Q. At low temperature we see  inelastic peaks on both sides of the central elastic peak at about 2.1 $\mu$eV. The inelastic peaks move towards the central elastic peak at higher temperatures and finally merge into the elastic peak at $T_C \approx 70$ K. We fitted three Gaussian peaks for the elastic and the two inelastic peaks by least squares method. We constrained the two inelastic peaks to have same widths. Fig. \ref{E(T)} shows the temperature variation of the energy of the inelastic peaks. The energy of the inelastic peak decreases continuously at first slowly then close to $T_C \approx 70$ K the energy becomes zero. The determination of the energy of the inelastic peak by fitting with Gaussian functions do not work when the inelastic peaks approach close to the central elastic peak close to $T_C$. The energy evaluated for $T = 75$ K is therefore not reliable. We have therefore drawn Fig. \ref{E(T)} a smooth curve passing through all other data points except that at T = 75 K. The dotted curve that gives  $T_C \approx 70$ K is just an extrapolation of the smooth curve.

Fig. \ref{tbspectra} shows Q-integrated neutron scattering spectra from Tb$_2$CoMnO$_6$ at several temperatures. At T = 2 K  we observe two clear inelastic signals at about E = 2.1 $\mu$eV on the energy-loss and energy-gain sides. At T = 20 K the inelastic signals are still visible. At T = 40 K, however, strong quasielastic scattering appears and presumably obliterates the inelastic signals. At T = 60 K the quasielastic scattering is even stronger and the inelastic signals are not visible. At T = 80 K the quasielastic scattering becomes very broad and most of the quasielastic scattering goes outside the window making the inelastic signal visible again. At T= 100 K there exist no appreciable inelastic signals. We know from magnetization measurements the transition temperature T$_c$ of Tb$_2$CoMnO$_6$ is also about 100 K. However, attempts to fit the inelastic signals at T = 2, 20 and 80 K were not very successful. The quasielastic scattering at T = 40 and 60 K could be fitted by a Lorentzian function and a Gaussian resolution function. We interpret the origin of quasielastic scattering in Tb$_2$CoMnO$_6$ to be due to the fluctuating electronic moment of Tb ions which is very large. In contrast we do not observe any quasielastic scattering in Y$_2$CoMnO$_6$ because of the absence of a magnetic ion in the rare earth site in this compound.  Also we know from magnetization measurements that the magnetic ordering in Tb$_2$CoMnO$_6$ is not quite ferromagnetic but is probably spin-glass like.

\section{Hyperfine interaction in Co-compounds}

  In Table 1 we give the magnetic moments and the energy of low energy nuclear spin excitations in Co and all Co compounds studied so far by high resolution neutron spectroscopy.  In Fig. \ref{cocompounds} we plot the energy of nuclear spin excitations vs. the ordered magnetic moments of  Co compounds investigated so far. We note that although Co, CoO, CoCl$_2$ and amorphous Co-P alloys and also perhaps LaCo$_{13}$ lie on a straight line passing through the origin, several other compounds viz. LaCo$_5$, YCo$_5$, ThCo$_5$ CoF$_2$, CoV$_2$O$_6$ and Co$_2$SiO$_4$ deviate appreciably from the linear behaviour. The slope of the linear fit $E = a\mu$ ($\mu$ = magnetic moment) of the
data for  Co, CoCl$_2$, Co-P amorphous alloys, CoO  and LaCo$_{13}$ gives a value of $a=0.50 \pm 0.01 \mu eV$/$\mu_B$. The data for these compounds have been shown by blue circles and the fitted blue straight line. The data corresponding to the other \emph{anomalous compounds} can be fitted by a power law $E=a\mu^n$ with $a = 0.19 \pm 0.03$ and $n= 1.5 \pm 0.1\approx 3/2$ and is shown by the red curve in Fig. \ref{cocompounds}. We ascribe the anomalous behaviour to the presence of unquenched orbital moments in these compounds. The orbital moment in 3d transition metal compounds is normally quenched. However Co and also V compounds are known to possess considerable unquenched orbital moments. The hyperfine field due to the unquenched orbital moment can have opposite sign \cite{freeman65} to that due to Fermi contact term and thus can reduce the effective hyperfine field. This is probably the case for the Co compounds, CoF$_2$, Co$_2$SiO$_4$, CoV$_2$O$_6$ and intermetallic compounds LaCo$_{13}$, LaCo$_5$, YCo$_5$ and ThCo$_5$. The hyperfine fields in these compounds are much less than that expected from their moments and therefore deviate from the linear behaviour. This is of course only a qualitative explanation in the absence of any ab-intitio calculations of the orbital moments and hyperfine fields in these compounds. Assuming the double perovskite compounds Y$_2$CoMnO$_6$ and Tb$_2$CoMnO$_6$ behave normally we can estimate the total magnetic moment from the straight line plot to be $4.2 \mu_B$ for both the compounds.  If however we assume that the double perovkite compounds behave anomalously like those lying on the red curve of Fig. \ref{cocompounds} then we get a magnetic moment of $4.8 \mu_B$. The estimated magnetic moments are very large compared to the expected spin-only value of 3.5 $\mu_B$. This simply shows that the hyperfine field or the hyperfine splitting is not any simple function of the magnetic moment. To clarify these points ab-intio calculations of the hyperfine field and magnetic moments of Co compounds are urgently needed. It is also probably helpful to investigate more Co compounds experimentally by the present technique.


\begin{thebibliography}{99}
\bibitem{wold58}A. Wold, R.J. Arnott, and J.B. Goodenough, J. Appl. Phys {\bf 29}, 387
\bibitem{goodenough61}J.B. Goodenough, A. Wold, R.J. Arnott, and N. Menyuk, Phys. Rev. B {\bf 124}, 373 (1961).
\bibitem{goodenough55}J.B. Goodenough, Phys. Rev. {\bf 100}, 564 (1955).
\bibitem{kanamori59}J. Kanamori, J. Phys. Chem. Solids {\bf 10}87 (1959).
\bibitem{rogado05}N. S. Rogado, J. Li, A. W. Sleight, M. A. Subramanian, Adv. Mater., 17 (2005) 2225; D. J. Singh, C. H.
Park, Phys. Rev. Lett., 100 (2008) 087601
\bibitem{bull03}C. L. Bull, D. Gleeson, K. S. Knight, J. Phys.: Condens. Matter, 15 (2003) 4927
\bibitem{ogale99}A. S. Ogale, S. B. Ogale, R. Ramesh, T. Venkatesan, Appl. Phys. Lett. 75 (1999) 537
\bibitem{booth09}R. J. Booth, R. Fillman, H. Whitaker, A. Nag, R. M. Tiwari, K. V. Ramanujachary, J. Gopalakrishnan, S.
E. Lofland, Mater. Res. Bull. 44 (2009) 1559
\bibitem{truong11} K. D. Truong, M. P. Singh, S. Jandl, P. Fournier, J. Phys.: Condens. Matter, {\bf 23},  052202 (2011)
\bibitem{sazanov07}A. P. Sazanov, I. O. Troyanchuk, M. Kopcewicz, V. V. Sikolenko, U. Zimmermann, K. B\"arner, J. Phys.:
Condens. Matter, 19 (2007) 046218
 \bibitem{freeman65}A.J. Freeman and R.E. Watson, in \emph{Magnetism} ed. G.T. Rado and H. Suhl, Academic Press, New York, London, vol. II A (1965).
 \bibitem{heidemann70} A. Heidemann, Z. Phys. {\bf 238}, 208 (1970).
   \bibitem{chatterji00} T. Chatterji and B. Frick, Physica B {\bf 276-278}, 252 
            (2000)
\bibitem{chatterji02} T. Chatterji and B. Frick, Appl. Phys. A 
            {\bf 74}[Suppl.], S652 (2002)
\bibitem{chatterji04} T. Chatterji and B. Frick, Physica B {\bf 350}, e111 (2004)
 \bibitem{chatterji04a}T. Chatterji and B. Frick, Solid State Comm. {\bf 131}, 453 (2004).
 \bibitem{chatterji08}T. Chatterji, G.J. Schneider and R.M. Galera, Phys. Rev. B {\bf 78}, 012411 (2008).
 \bibitem{chatterji08a}T. Chatterji, G.J. Schneider, L. van Eijk, B. Frick and D. Bhattacharya, J. Phys: Condens. Matter {\bf 21}, 126003 (2009).
\bibitem{chatterji09f}T. Chatterji and G.J. Schneider, Phys. Rev. B {\bf 9}, 132408 (2009).
   \bibitem{chatterji09c}T. Chatterji and G.J. Schneider, Phys. Rev. B  {\bf 79}, 212409 (2009).
    \bibitem{chatterji09d}T. Chatterji and G.J. Schneider, J. Phys.: Condens. Matter {\bf 21}, 436008 (2009).
 \bibitem{chatterji10}T. Chatterji, J. Wuttke and A.P. Sazonov, J. Magn. Magn. Mater {\bf 322}, 3148(2010).
 \bibitem{chatterji11}T. Chatterji, J. Wuttke and S.A.J. Kimber,  unpublished results (2011)
 \bibitem{chatterji10a}T. Chatterji and J. Wuttke (unpublished results).
 \bibitem{chatterji11a}T. Chatterji and B. Frick (unpublished results)
 \bibitem{heidemann75a} A. Heidemann, Z. Phys. B {\bf 20}, 385 (1975)
\bibitem{heidemann75b} A. Heidemann, D. Richter and K.H.J. Buschow, Z. Phys. B {\bf 22}, 367 (1975).
\end{thebibliography}
\end{document}